\documentstyle[12pt,dina4p,amssym,epsfig]{article}

\newcommand{\hdick}{\noalign{\hrule height1.4pt}}
\newcommand{\GeV}{\rm GeV}

\begin{document}
\thispagestyle{empty}


\noindent
{\large DESY 95-079,  hep-ex/yymmnn \\[.2em]
       April 1995}

\vspace{5cm}
\begin{center}
  {\Large \bf
    Leptoquarks and Compositeness Scales \\[.15em]
    from a Contact Interaction Analysis of \\[.3em]
    Deep Inelastic $e^\pm \, p$ Scattering at HERA} \\[2cm]
  {\Large H1 Collaboration}
\end{center}

\vspace{10cm}
\begin{center}
  {\large ISSN 0418-9833}
\end{center}


\newpage
\thispagestyle{empty}

\setcounter{page}{1}
\noindent
DESY 95-079 \hfill ISSN 0418-9833 \\
April 1995

\vspace{4cm}
\begin{center}
  {\large \bf
    Leptoquarks and Compositeness Scales from a Contact Interaction \\[.3em]
    Analysis of Deep Inelastic $e^\pm \, p$ Scattering at HERA} \\[1.5em]
  {\large H1 Collaboration}
\end{center}

\vspace{1cm}

\begin{quotation}

\noindent
{\bf Abstract.}
A contact interaction analysis is presented to search for new phenomena beyond
the Standard Model in deep inelastic
$e^\pm p \rightarrow e^\pm \, hadrons$ scattering.
The data are collected with the H1 detector at HERA and correspond to
integrated
luminosities of $0.909 \ {\rm pb}^{-1}$ and $2.947 \ {\rm pb}^{-1}$ for
electron
and positron beams, respectively.
The differential cross sections $d\sigma / dQ^2$ are measured in the $Q^2$
range
bet\-ween $160 \ \GeV^2$ and $20,000 \ \GeV^2$.
The absence of any significant deviation from the Standard Model prediction is
used to constrain the couplings and masses of new leptoquarks
and to set limits on electron--quark compositeness scales
and on the radius of light quarks.

\end{quotation}

\newpage

\begin{flushleft}
 S.~Aid$^{13}$,                   
 V.~Andreev$^{25}$,               
 B.~Andrieu$^{28}$,               
 R.-D.~Appuhn$^{11}$,             
 M.~Arpagaus$^{36}$,              
 A.~Babaev$^{24}$,                
 J.~B\"ahr$^{35}$,                
 J.~B\'an$^{17}$,                 
 Y.~Ban$^{27}$,                   
 P.~Baranov$^{25}$
% ...

\bigskip\bigskip
{\it
% ...
$ ^{11}$ DESY, Hamburg, Germany$^a$ \\
% ...
$ ^{13}$ II. Institut f\"ur Experimentalphysik, Universit\"at Hamburg,
                                     Hamburg, Germany$^ a$  \\
% ...
$ ^{17}$ Institute of Experimental Physics, Slovak Academy of
                Sciences, Ko\v{s}ice, Slovak Republic$^ f$\\
% ...
$ ^{24}$ Institute for Theoretical and Experimental Physics,
                                                 Moscow, Russia \\
 $ ^{25}$ Lebedev Physical Institute, Moscow, Russia$^ f$ \\
% ...
$ ^{27}$ LAL, Universit\'{e} de Paris-Sud, IN2P3-CNRS,
                            Orsay, France\\
 $ ^{28}$ LPNHE, Ecole Polytechnique, IN2P3-CNRS,
                             Palaiseau, France \\
% ...
$ ^{35}$ DESY, Institut f\"ur Hochenergiephysik,
                              Zeuthen, Germany$^ a$\\
 $ ^{36}$ Institut f\"ur Teilchenphysik,
          ETH, Z\"urich, Switzerland$^ i$\\
% ...
$ ^a$ Supported by the Bundesministerium f\"ur
                                  Forschung und Technologie, FRG
 under contract numbers 6AC17P, 6AC47P, 6DO57I, 6HH17P, 6HH27I, 6HD17I,
 6HD27I, 6KI17P, 6MP17I, and 6WT87P \\
% ...
$ ^f$ Supported by the Deutsche Forschungsgemeinschaft\\
% ...
$ ^i$ Supported by the Swiss National Science Foundation\\
   } 
\end{flushleft}

\newpage

\section{Introduction}

The $e^\pm\, p$ collider HERA, in which $27.5 \ \GeV$ ($26.7 \ \GeV$ in 1993)
leptons collide with
$820 \ \GeV$ protons, provides access to an as yet unexplored mass domain
for the discovery of new particles.
Several extensions of the Standard Model postulate either new fermions
or new bosons, such as leptoquarks and additional gauge bosons.
Signals for physics beyond the Standard Model can be discovered either directly
or indirectly.

Common to all {\em direct} searches is the $s$ channel formation of a new heavy
resonance at a mass $M_X = \sqrt{x\,s}$, which is, however, limited by the
available centre of mass energy of $\sqrt{s} \simeq 300 \ \GeV$.
The scaling variable $x$ is the momentum fraction of
the proton carried by the struck quark.
Results of direct searches for new heavy bosons at HERA have been published
recently by the experiments H1~\cite{h1lq} and ZEUS~\cite{zeuslq}.

The search for new bosons or $e\,q$ compositeness can be considerably extended
beyond the kinematic production limit through the study of {\em indirect}
effects from virtual particle exchange.
Such effects may become observable as deviations from the Standard
Model expectation at high momentum transfers $Q^2$.

The analysis presented here combines all data collected with the H1 experiment
at HERA during 1993 and 1994 with electron and positron beams.
They represent an increase in integrated luminosity by a factor of $\sim 9$
compared to the previous data sample~\cite{h1lq}.

\section{Phenomenology of $(\bar{e}\,e)\,(\bar{q}\,q)$ Contact Interactions}

New currents or bosons may produce indirect effects
through the interference of a {\em virtual} particle exchange with
the $\gamma$ and $Z$ fields of the Standard Model.
For particle masses well above the  available production energy,
such indirect signatures may be investigated by adding
general contact interaction terms to the Standard Model Lagrangian.
Sufficiently heavy particles $X$ cease to propagate and thus
new contact terms and modified vertices arise from `contracting' the particle
propagators to an effective 4--fermion point--like interaction.
The separate dependence of $s$, $t$ and $u$ channel amplitudes on couplings
$g_{X\rightarrow i,\;f}$ to states $i,\,f$ and mass $M_X$ reduce to the
dependence on effective couplings with dimension $[{\rm mass}^{-2}]$
\begin{eqnarray*}
 \eta_{if}&\equiv&\frac{g_{X\rightarrow i}\;
 g_{X\rightarrow f}}{M_X^2} \; .
\end  {eqnarray*}

The most general chiral invariant neutral current contact interaction
Lagrangian can be written in the form~\cite{haberl}
\begin{eqnarray*}
  {\cal L}^{NC}_{\rm contact}
               &=&\sum_{q \, = \, u,\, d}\left\{\eta^q_{LL}\,
   (\bar{e}_L\gamma_\mu e_L)(\bar{q}_L\gamma^\mu q_L)
   +\eta^q_{LR}\,
   (\bar{e}_L\gamma_\mu e_L)(\bar{q}_R\gamma^\mu q_R)\right.\\
   &&\ \ \ \ \left.+\;\eta^q_{RL}\,(\bar{e}_R\gamma_\mu e_R)
   (\bar{q}_L\gamma^\mu q_L) +\eta^q_{RR}\,
   (\bar{e}_R\gamma_\mu e_R)(\bar{q}_R\gamma^\mu q_R)\right\} \; ,
\end{eqnarray*}
where the indices $L$ and $R$ denote the left--handed and right--handed
fermion helicities and the sum extends over {\em up} and
{\em down} quarks and antiquarks $q$.

Although contact interactions have been originally proposed
in the context of composite leptons and quarks~\cite{eichten,rueckl}, this
ansatz can be easily applied to other new phenomena~\cite{haberl}
by an appropriate choice of the coupling coefficients $\eta_{if}$.

{\bf Leptoquarks} are colour triplet bosons of spin 0 or 1, carrying
lepton ($L$) and baryon ($B$) number and fractional electric charge.
They couple to lepton--quark pairs and appear in almost all extensions of the
Standard Model which try to establish a connection between leptons
and quarks~\cite{buchmueller,bschrempp}.
Leptons and quarks may be either arranged in common multiplets, like in Grand
Unified Theories or superstring motivated $E_6$ models, or they may have
a common substructure as in composite models.
A fermion number $F = L + 3\,B$ is defined, which takes the values
$F = 2$ for leptoquarks coupling to $e^-\,q$ and
$F = 0$ for leptoquarks coupling to $e^-\,\bar{q}$.
For positrons the fermion number $F$ changes by two units.
Consequently $F = 2$ leptoquarks are easier accessible in $e^-\,p$ scattering,
while positron beams are more sensitive to $F = 0$ leptoquarks,
since at moderate $x$ values quarks are more abundant in the proton than
antiquarks.

The notation, the contact interaction coefficients $\eta_{if}$ and the fermion
number assignment for leptoquarks with mass $M_{LQ}$ and coupling $\lambda$ are
given in Table~\ref{lqres} (from ref.~\cite{haberl}).
The only unknown is the ratio $M_{LQ}/\lambda$.
Note that vector leptoquarks have positive coupling coefficients, while scalar
leptoquarks have negative coupling coefficients, being a factor of 2 smaller in
magnitude.

In the Standard Model the fundamental particles -- leptons, quarks and gauge
bosons -- are assumed to be pointlike.
A possible fermion {\bf compositeness} or substructure can be expressed through
$ \eta_{if} \equiv \pm \, g^2 / \Lambda^{\pm \ 2}_{i f}$,
where the signs indicate posi\-tive and negative interference with the Standard
Model currents, $g$ is the coupling strength conventionally choosen as
$g^2/4\,\pi = 1$ and $\Lambda$ is the compositeness scale.

\section{The H1 Detector}

A detailed description of the H1 detector can be found
elsewhere~\cite{h1detector}.
Here only those components are described, which are relevant for the present
analysis.

The lepton energy and angle is measured in a finely segmented liquid
argon (LAr) sampling calorimeter covering the polar
angle\footnote{The incoming proton moves in the $+z$
      direction with polar angle $\theta=0^{\circ}$.}
range 4$^{\circ} \le \theta \le$ 153$^{\circ}$ and all azimuthal
angles.
It consists of a lead/argon electromagnetic section with a
thickness varying between 20 and 30 radiation lengths
and a stainless steel/argon section
for the measurement of hadronic energy flow, which offers in total a
containment varying from 4.5 up to 8 interaction lengths.
Electron energies are measured with a resolution
of $\sigma(E)/E \simeq$ $12$ \%/$\sqrt{E}\oplus1\%$ and hadron energies
with $\sigma(E)/E \simeq$ $50$ \%/$\sqrt{E}\oplus2\%$.
The absolute energy scales are known to 3\% and 5\% for
electrons and hadrons, respectively.
The angular resolution of the scattered lepton measured from the
electromagnetic shower in the calorimeter is $\sim 7 \ {\rm mrad}$.
A lead/scintillator electromagnetic backward calorimeter extends the
coverage at large angles
(155$^{\circ} \le \theta \le$ 176$^{\circ}$).
The instrumented iron flux return yoke is used to
measure the leakage of hadronic showers.

Located inside the calorimeters is a tracking system, which consists of
central drift and proportional chambers
        (25$^{\circ} \le \theta \le$ 155$^{\circ}$),
a forward track detector
        (7$^{\circ} \le \theta \le$ 25$^{\circ}$)
and backward proportional chambers
        (155$^{\circ} \le \theta \le$ 175$^{\circ}$).
The tracking chambers and calorimeters are surrounded
by a superconducting solenoid coil providing a uniform field of
1.15 T within the tracking volume.

The luminosity is determined from the rate of the Bethe--Heitler process
$e\, p \rightarrow e\, p\, \gamma$ measured in a luminosity
monitor~\cite{h1lumi} far downstream the electron direction.
The systematic errors of the integrated luminosity vary between
1.8\% (1994 data) and 4.5\% (1993 data).

\section{Data Selection and Analysis}

The analysis is based on a purely inclusive measurement of the final state
lepton in deep inelastic neutral current events
$e^\pm\,p \rightarrow e^\pm \ hadrons$.
All quantities of event properties are determined from the calorimeters alone,
tracking information is only used to get the primary vertex position.
The kinematic variables $Q^2$, the negative squared momentum transfer, and the
scaling variable $y$ are derived from the scattered lepton energy $E'_e$
and polar angle $\theta_e$
\begin{eqnarray*}
  Q^2 & = & 4\,E_e\,E'_e\,\cos^2\frac{\theta_e}{2} \ , \\
  y   & = & 1 - \frac{E'_e}{E_e}\,\sin^2\frac{\theta_e}{2} \ ,
\end{eqnarray*}
where $E_e$ is the lepton beam energy.
These quantities are related to the Bjorken scaling variable $x$
by $Q^2 = x\,y\,s$.

The data were taken in three periods:
(i)   during 1993 with $26.7\ \GeV$ electrons,
(ii)  during 1994 with $27.5\ \GeV$ electrons, and
(iii) during 1994 with $27.5\ \GeV$ positrons.
The event selection is similar to the previous leptoquark
analysis~\cite{h1lq} with the following requirements:
\begin{enumerate}
\item The transverse energy of the scattered lepton has to exceed
      $E'_{\perp,\,e} > 8\ \GeV$.  
\item The polar angle of the scattered lepton has to be within the LAr
      calorimeter acceptance
      $10^\circ < \theta_e < 150^\circ$.
\item A primary vertex has to be reconstructed within
      $ |\, z_{vertex} - \langle z \rangle \, | < 35 \ {\rm cm}$
      of the nominal interaction point $\langle z \rangle$.
\item The energy--longitudinal momentum conservation
      $|\, \sum \ (E - p_z) - 2\,E_e \, | < 10 \ \GeV$
      must be satisfied, where the sum extends over all detected particles or
      energy clusters.
\item The event has to be balanced in transverse momentum
      $|\, \vec{p}_\perp^{\ evt} \, | < 15 \ \GeV$.
\item The scaling variable $y$ has to fullfil $ y < 0.8 $.
\end{enumerate}

Requirements (1) and (2) assure that the kinematic quantities are well
measured in the LAr calorimeter. $Q^2$ is always well measured, while the
resolution in $y$ degrades at low values of $y \lesssim 0.1$.
The requirements (3) and (5) suppress beam--wall and beam--gas background.
The criteria (4) and (5) provide a good containment of the final state
particles
and reject events with a hard photon radiated from the
initial state lepton and photoproduction events with a misidentified lepton in
the LAr calorimeter.
For events with all final state particles detected (except the proton
remnants) one expects $\sum \ (E - p_z) \approx 2\,E_e$.
Requirement (6) is introduced to
avoid the region affected by large radiative corrections
and to suppress photoproduction events with a misidentified lepton.
The remaining contamination from photoproduction, beam--gas collisions and
cosmic rays is negligible.

The final data samples consist of
(i)   739 events for an integrated luminosity of
      ${\cal L} = 0.418 \ (\pm 4.5 \%) \ {\rm pb}^{-1}$
      of the $e^-\,p$ 1993 data,
(ii)  810 events for ${\cal L} = 0.491 \ (\pm 2.4 \%) \ {\rm pb}^{-1}$
      of the $e^-\,p$ 1994 data, and
(iii) 5201 events for ${\cal L} = 2.947 \ (\pm 1.8 \%) \ {\rm pb}^{-1}$
      of the $e^+\,p$ 1994 data.

The measured cross sections $d\sigma / dQ^2$ are corrected for detector effects
and QED radiation and extrapolated to the full kinematic phase space.
The correction factors for each $Q^2$ bin are obtained from Monte Carlo
studies.
Deep inelastic scattering events are generated according to the Standard Model
cross section
\begin{eqnarray*}
      \frac{d^2\sigma(e^\pm \, p \rightarrow e^\pm \, X)}{dx\,dQ^2} & = &
        \frac{2\,\pi\,\alpha^2}{x\,Q^4}\,
        \left\{ \, Y_+\,F_2(x,Q^2)
        -y^2\,F_L(x,Q^2)
        \mp Y_-\,x\,F_3(x,Q^2) \, \right\}\ ,  \\
        Y_\pm & = & 1 \pm (1 - y)^2
\end{eqnarray*}
using the parton densities from the MRS~H
parametrization~\cite{mrs} of the structure functions $F_i(x,Q^2)$.
The longitudinal structure function $F_L$ has not yet been measured at HERA,
but is expected to give very small contributions only at low $Q^2$ and low $x$
values.
For the present analysis $F_L$ has been neglected.
Radiative effects are taken into account by
the event gene\-rator DJANGO~6~\cite{django},
which includes the $\cal{O}(\alpha)$
electroweak corrections and the QCD matrix elements to first order in
$\alpha_s$, supplemented by leading--logarithmic parton showers.
The lepton detection and measurement is simulated by smearing the event vertex
and the generated four-vector according to measured resolutions, acceptances
and
efficiencies, as determined from data.
The hadron final state is simulated by smearing the four-vector of the `struck
quark' as calculated from the electron kinematics (no hadronization).
This very fast and efficient acceptance simulation
is completely adequate to describe all properties
of the deep inelastic scattering events used in the present
inclusive analysis.


\section{Results}

\begin{figure}[p]
\vspace{5cm}
  \begin{center}
    \hspace*{-5.5cm} \vspace*{-.0cm}
    \unitlength 1mm
    \begin{picture}(120,120)
    \put(5,0) {\mbox{\epsfig{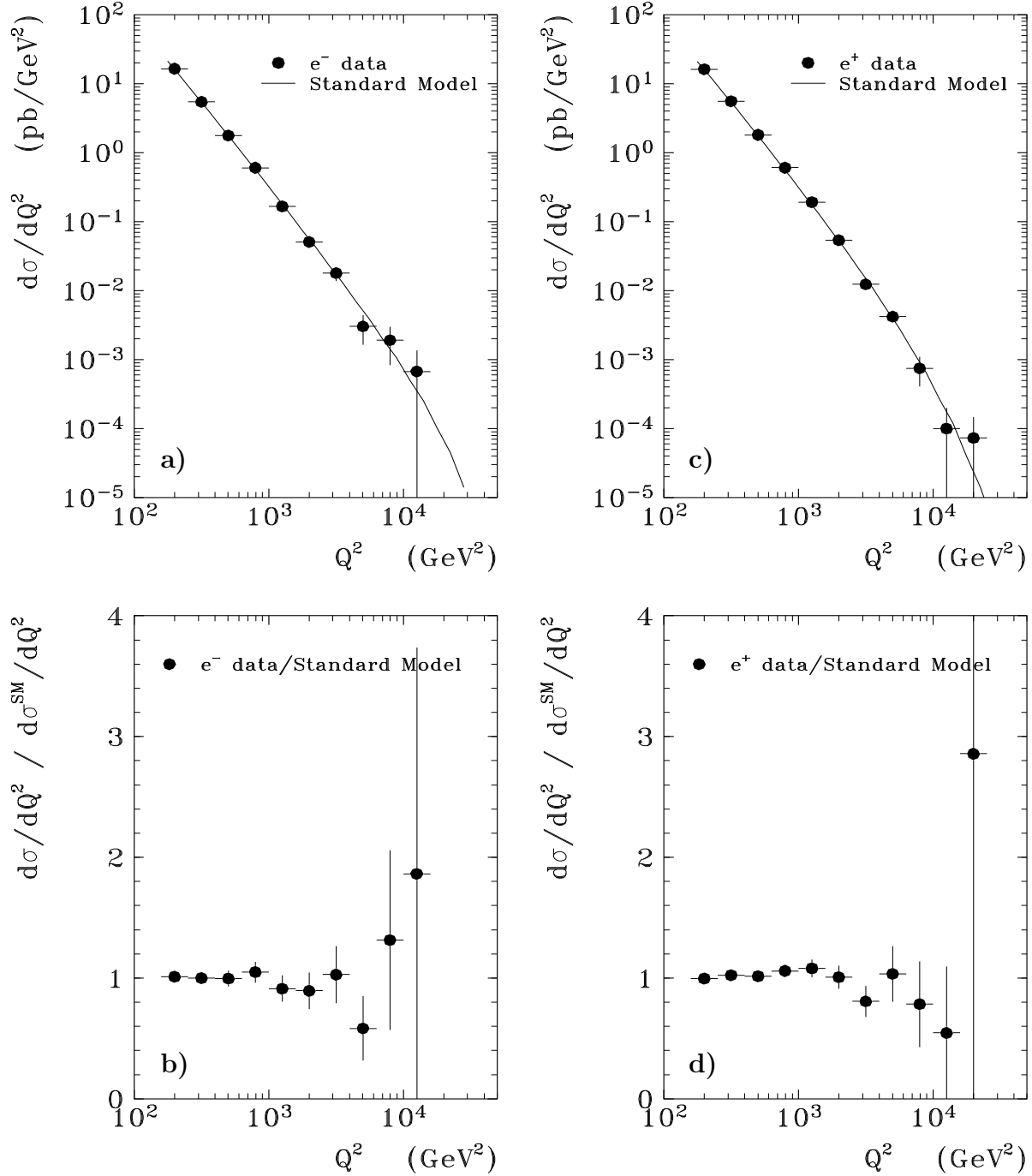}}}
    \put(30,112) {\bf a)}
    \put(112,112) {\bf c)}
    \put(30, 18) {\bf b)}
    \put(112, 18) {\bf d)}
  \end{picture}
  \end{center}
  \caption{Differential cross sections $d\sigma / dQ^2$ versus $Q^2$
           ($\bullet$)
           in comparison with the Standard Model expectations (---).
           {\bf a)} and {\bf b)}  $e^-\,p$ data,
           {\bf c)} and {\bf d)}  $e^+\,p$ data.
           The error bars represent statistical and systematic errors added
           in quadrature, except for an overall normalization
           uncertainty of $3.5\%$ ($e^-\,p$ data) and
           $1.8\%$ ($e^+\,p$ data), respectively.}
  \label{xsection}
\end{figure}

{\bf Cross Sections} \\
The corrected differential cross sections $d\sigma / dQ^2$ for the $e^-\,p$
and $e^+\,p$ data are shown in Fig.~\ref{xsection} and
listed in Table~\ref{tabxsection}.
Statistical and systematic errors are added in quadrature, except for an
overall normalization uncertainty.
The systematic errors include uncertainties of the trigger and detection
efficiencies, the vertex reconstruction
and the lepton energy calibration of the LAr calorimeter
(all determined from the data),
as well as uncertainties due to radiative corrections and the choice of
different parton distributions (MRS~H, MRS~D$^0$' and MRS~D$^-$').
The syste\-matic uncertainties are typically $\sim 2\%$, except for the energy
calibration, which contributes $\sim 6\%$ in the low $Q^2$ region.
The overall normalization errors are $3.5\%$ for the $e^-\,p$ data and
$1.8\%$ for the $e^+\,p$ data and account for uncertainties of the luminosity
measurement.
The acceptance is a smooth function of the squared momentum transfer.
It rises from $\sim 65 \%$ at the lowest $Q^2$
to $\sim 80\%$ at $Q^2 \simeq 1000 \ \GeV^2$ and then slowly
decreases to $\sim 60\%$ at $Q^2 \simeq 10,000 \ \GeV^2$.
The extrapolation to the full phase space is not critical for the low and
medium $Q^2$ region, where the validity of the Standard Model is well
established.
At high $Q^2$ and large $y$ the Standard Model is assumed.
The effects of the extrapolation on the contact interaction analysis
are, however, small.

\begin{table}[htb]
\begin{center}
  \begin{tabular}{ c c c }
    \hdick \\[-.5em]
       & \ $e^-\,p$ data \ & \ $e^+\,p$ data \ \\[.2em]
    \ $\langle Q^2 \rangle \ \ [ \GeV^2 ]$ \ &
    \ \quad\quad $d\sigma / dQ^2 \ \ [ {\rm pb} \ \GeV^{-2} ]$ \ \quad\quad &
    \ \quad\quad $d\sigma / dQ^2 \ \ [ {\rm pb} \ \GeV^{-2} ]$ \ \quad\quad
    \\[.5em]
    \hdick \\[-.5em]
   \ $   200 $ \
 & \ $ 16.5 \pm   0.7 \pm   0.6$ \
 & \ $ 16.2 \pm   0.4 \pm   0.8$ \ \\
   \ $   316 $ \
 & \ $ 5.43 \pm  0.25 \pm  0.11$ \
 & \ $ 5.57 \pm  0.14 \pm  0.16$ \ \\
   \ $   501 $ \
 & \ $ 1.77 \pm  0.11 \pm  0.04$ \
 & \ $ 1.81 \pm  0.06 \pm  0.06$ \ \\
   \ $   794 $ \
 & \ $( 6.03 \pm  0.48 \pm  0.11 )\cdot 10^{-1}$ \
 & \ $( 6.04 \pm  0.27 \pm  0.15 )\cdot 10^{-1}$ \ \\
   \ $  1259 $ \
 & \ $( 1.66 \pm  0.19 \pm  0.05 )\cdot 10^{-1}$ \
 & \ $( 1.92 \pm  0.12 \pm  0.06 )\cdot 10^{-1}$ \ \\
   \ $  1995 $ \
 & \ $( 5.09 \pm  0.84 \pm  0.21 )\cdot 10^{-2}$ \
 & \ $( 5.38 \pm  0.49 \pm  0.13 )\cdot 10^{-2}$ \ \\
   \ $  3162 $ \
 & \ $( 1.79 \pm  0.40 \pm  0.09 )\cdot 10^{-2}$ \
 & \ $( 1.23 \pm  0.19 \pm  0.04 )\cdot 10^{-2}$ \ \\
   \ $  5012 $ \
 & \ $( 3.02 \pm  1.14 \pm  0.77 )\cdot 10^{-3}$ \
 & \ $( 4.19 \pm  0.91 \pm  0.18 )\cdot 10^{-3}$ \ \\
   \ $  7943 $ \
 & \ $( 1.91 \pm  0.95 \pm  0.51 )\cdot 10^{-3}$ \
 & \ $( 7.50 \pm  3.36 \pm  0.48 )\cdot 10^{-4}$ \ \\
   \ $ 12589 $ \
 & \ $( 6.74 \pm  6.74 \pm  0.66 )\cdot 10^{-4}$ \
 & \ $( 1.00 \pm  1.00 \pm  0.11 )\cdot 10^{-4}$ \ \\
   \ $ 19953 $ \
 & \ ---
 & \ $( 7.32 \pm  7.32 \pm  1.49 )\cdot 10^{-5}$ \ \\
  \end{tabular}
\end{center}
\caption{Differential cross sections $d\sigma / dQ^2$ for the
         $e^-\,p$ and $e^+\,p$ data with statistical and systematic
         errors.
         There is an additional overall normalization
         uncertainty of $3.5\%$ ($e^-\,p$ data) and
         $1.8\%$ ($e^+\,p$ data), respectively,
         due to systematic errors of the luminosity measurement.}
\label{tabxsection}
\end{table}

The measured cross sections are well described by the Standard Model
expectations over five orders of magnitude in the $Q^2$ range between
$160 \ \GeV^2$ and $20,000 \ \GeV^2$.
No significant deviation is observed for either lepton charge,
see Fig.~\ref{xsection}.

It is interesting to note that the $e^-\,p$ cross section tends to be
slightly higher than the $e^+\,p$ cross section at high $Q^2$ values, as
expected from the different couplings of the leptons to the $Z$ boson.
The charge asymmetry is, however, not yet significant due to the limited
electron data statistics.

\bigskip
The contact interaction analysis
investigates the differential cross sections $d\sigma / dQ^2$ and
interpretes any deviation from the Standard Model as lower limits
on the ratio mass over coupling strength of new heavy leptoquarks
or composite $e\,q$ structures.
It is assumed that possible deviations are caused by only one new boson
exchange at the time.
A combined $\chi^2$ analysis of the $e^-\,p$ and $e^+\,p$ data is
performed including for each data set
an individual overall normalization constant $f_{norm}$ with its corresponding
error $\Delta_{norm}$ (see discussion above)
\begin{eqnarray*}
\chi^2 & = & \sum_{l\, =\, e^-, \, e^+} \left \{ \
             \sum_{k} \, \left (\frac{\sigma_k^l\,(Q^2)\, f^l_{norm}
             - \sigma^{theor}(Q^2,\eta_{if})}
           {\Delta \sigma_k^l\,(Q^2) \, f^l_{norm}}\right )^2
           +\left(\frac{f^l_{norm}-1}{\Delta^l_{norm}} \right )^2
           \ \right \} \ .
\end{eqnarray*}
The inner sum extends over the bins of one data set, while the outer sum is
taken over both the $e^-\,p$ and $e^+\,p$ data.
Thus, fitted parameters are the $\eta_{if}$ coefficients of the contact
interaction model and the two normalization constants.
Limits at 95\% confidence level are derived from the increase of $\chi^2$ by
3.89 with respect to its minimum value, which in most cases coincides with the
Standard Model fit.
Enlarging the normalization errors arbitrarily by $2 \%$ would lower
the resulting limits by $\sim 5 \%$.
Choosing different parton distributions, MRS~D$^0$' or MRS~D$^-$',
in the cross section calculation changes the limits by $\sim 3 \%$ in either
direction.

\bigskip\noindent
{\bf Leptoquarks} \\
The results of the leptoquark analysis are summarized in Table~\ref{lqres}.
Only those lower limits on $M_{LQ}/\lambda$ are quoted, which exceed the
kinematic phase space of HERA for direct production assuming a strong coupling
of $\lambda = 1$.

\begin{table}[htb]
\begin{center}
\begin{tabular}{c c c c c}
  \hdick \\[-1.5ex]
  \ leptoquark \ & \ coupling to $u$ quark \ & \ coupling to $d$ quark \
    & \ $F$ \ & \ $M_{LQ}/\lambda$ \ \\[.5ex]
    & $[ \GeV^{-2} ]$ & $[ \GeV^{-2} ]$ & & $[ \GeV ]$ \\[1ex]
  \hdick \\[-1.5ex]
  \ $S_0^L$ &
    \ $\eta^u_{LL} = -\frac{1}{2}\ (\lambda/M_{LQ})^2$ \ & & \ 2 \ \\
  \ $S_0^R$ &
    \ $\eta^u_{RR} = -\frac{1}{2}\ (\lambda/M_{LQ})^2$ \ & & \ 2 \ \\
  \ $\tilde{S}_0^R$ & &
    \ $\eta^d_{RR} = -\frac{1}{2}\ (\lambda/M_{LQ})^2$ \   & \ 2 \
    & \ $ 350$ \ \\
  \ $S_{1/2}^L$ &
    \ $\eta^u_{LR} = -\frac{1}{2}\ (\lambda/M_{LQ})^2$ \ & & \ 0 \ \\
  \ $S_{1/2}^R$ &
    \ $\eta^u_{RL} = -\frac{1}{2}\ (\lambda/M_{LQ})^2$ \  &
    \ $\eta^d_{RL} = -\frac{1}{2}\ (\lambda/M_{LQ})^2$ \  & \ 0 \  \\
  \ $\tilde{S}_{1/2}^L$ & &
    \ $\eta^d_{LR} = -\frac{1}{2}\ (\lambda/M_{LQ})^2$ \  & \ 0 \
    & \ $ 360$ \ \\
  \ $S_1^L$ &
    \ $\eta^u_{LL} = -\frac{1}{2}\ (\lambda/M_{LQ})^2$ \  &
    \ $\eta^d_{LL} = -1\ (\lambda/M_{LQ})^2$              & \ 2 \
    & \ $ 340$ \ \\[1ex]
  \hline \\[-1.5ex]
  \ $V_0^L$ & &
    \ $\eta^d_{LL} = +1\ (\lambda/M_{LQ})^2$ \   & \ 0 \  \\
  \ $V_0^R$ & &
    \ $\eta^d_{RR} = +1\ (\lambda/M_{LQ})^2$ \   & \ 0 \  \\
  \ $\tilde{V}_0^R$ &
    \ $\eta^u_{RR} = +1\ (\lambda/M_{LQ})^2$ \ & & \ 0 \
    & \ $ 760$ \ \\
  \ $V_{1/2}^L$ & &
    \ $\eta^d_{LR} = +1\ (\lambda/M_{LQ})^2$ \   & \ 2 \
    & \ $ 300$ \ \\
  \ $V_{1/2}^R$ &
    \ $\eta^u_{RL} = +1\ (\lambda/M_{LQ})^2$ \  &
    \ $\eta^d_{RL} = +1\ (\lambda/M_{LQ})^2$ \   & \ 2 \
    & \ $ 710$ \ \\
  \ $\tilde{V}_{1/2}^L$ &
    \ $\eta^u_{LR} = +1\ (\lambda/M_{LQ})^2$ \ & & \ 2 \
    & \ $ 800$ \ \\
  \ $V_1^L$ &
    \ $\eta^u_{LL} = +2\ (\lambda/M_{LQ})^2$ \  &
    \ $\eta^d_{LL} = +1\ (\lambda/M_{LQ})^2$ \   & \ 0 \
    & \ $ 1020$ \ \\[1ex]
  \hline
\end{tabular}
\end{center}
\caption[dum] {Contact interaction coefficients $\eta^q_{if}$,
               fermion number $F$ 
               and lower limits at 95\% confidence level on
               $M_{LQ}/\lambda$
               for scalar (S) and vector (V) leptoquarks.
               The leptoquark notation indicates the lepton chirality
               {\em L, R} and the weak isospin $I = 0,\ 1/2,\ 1$.
               The leptoquarks $\tilde{S}$ and $\tilde{V}$ differ by
               two units of hypercharge from $S$ and $V$, respectively.}
\label{lqres}
\end{table}

One notices that vector leptoquarks with a positive coupling to $u$ quarks
provide the most restrictive bounds approaching ${\cal O}\,(1\ {\rm TeV})$.
The sensitivity to scalar leptoquarks is generally lower by a factor of $\sim
2$
and two of them, $\tilde{S}_0^R$ and $\tilde{S}_{1/2}^L$, have only couplings
to
$d$ quarks.
It is not obvious which leptoquarks will provide the most stringent
limits, when arguing alone on the basis of the assigned fermion numbers $F$
and quark densities in the proton.
In addition to the $s$ channel amplitudes of direct production the contact
interaction ansatz implicitely contains the crossed $t$ and $u$ channel
diagrams.
Moreover, not all coupling coefficients $\eta_{if}$ contribute with the same
weight, thus spoiling the naive expectation\footnote{A comprehensive general
  discussion on the sensitivity of the coupling coefficients in the
  eight-dimensional $\eta_{if}$ space can be found in ref.~\cite{haberl}.}.

\begin{figure}[htb]
\vspace{-4.7cm}
  \begin{center}
    \hspace*{-5.5cm} \vspace*{-.6cm}
    \unitlength 1mm
    \begin{picture}(120,120)
    \put(5,0)  {\mbox{\epsfig{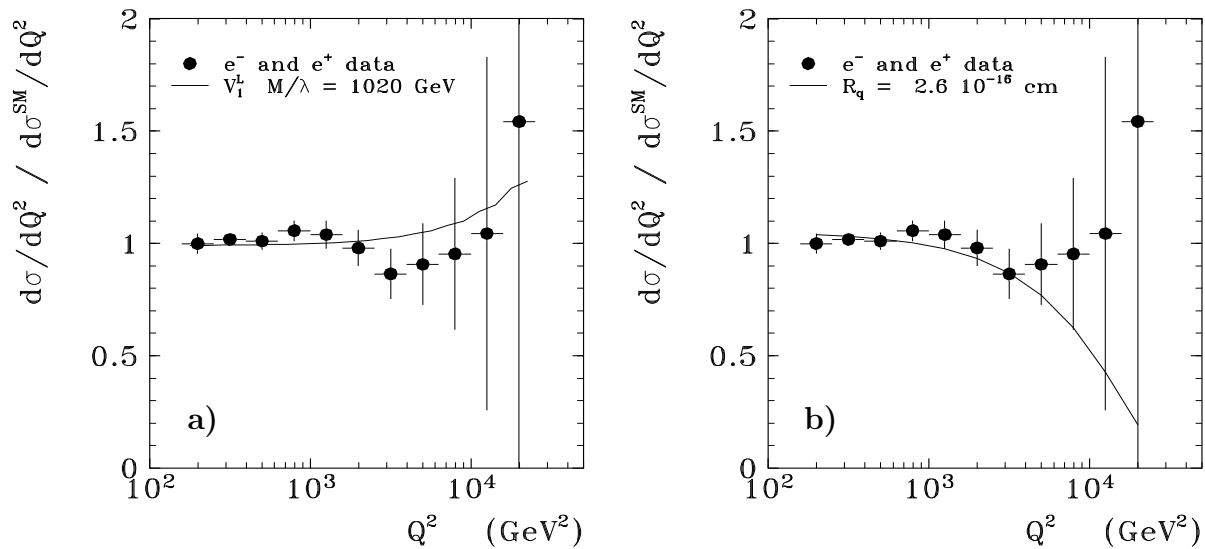}}}
    \put(30, 18) {\bf a)}
    \put(112, 18) {\bf b)}
  \end{picture}
  \end{center}
  \caption{Differential cross sections $d\sigma / dQ^2$
           normalized to the Standard Model expectation versus $Q^2$
           for the combined $e^-\,p$ and $e^+\,p$ data ($\bullet$).
           The error bars represent statistical and systematic errors added
           in quadrature, but do not contain an overall luminosity
           uncertainty (see text).
           {\bf a)} The curve shows the allowed contribution
           (95\% confidence level)
           of a vector leptoquark $V_1^L$
           with $ M_{LQ}/\lambda = 1020 \ \GeV$.
           {\bf b)} The curve shows the effect of a form factor
           with a quark radius $R_q = 2.6 \cdot 10^{-16}\ {\rm cm}$
           (95\% confidence level).}
  \label{xratio}
\end{figure}

As an illustration of the sensitivity of the data to virtual leptoquark
exchange
Fig.~\ref{xratio}~a shows the allowed contribution of a vector leptoquark
$V_1^L$.
The contact interaction contributions to the Standard Model rise with $Q^2$
as expected and amount to $\sim 25\%$ at the highest $Q^2$ values.

The {\em indirect} limits $M_{LQ}/\lambda$ derived from virtual leptoquark
exchange nicely complement the {\em direct} searches~\cite{h1lq,zeuslq},
because they exceed the kinematic range of HERA at large couplings.
The couplings $\lambda$ can be safely extracted down to
the nominal centre of mass energy,
since the present integrated luminosities restrict the accessible masses
to $M_{LQ} = \sqrt{x\,s} \, \lesssim \, 250 \ \GeV$.
For example, a vector leptoquark $V_1^L$ with a mass of
$300 \ \GeV$ can be excluded for couplings larger than the
electromagnetic strength $\lambda > \sqrt{4\,\pi\,\alpha} = 0.3$.

The limits given in Table~\ref{lqres} can be compared to those derived from
other, primarily low energy experiments.
The strongest bounds for leptoquarks coupling to the first lepton and quark
generations arise from atomic parity violation experiments and from
universality in leptonic $\pi$ decays.
Davidson {\em et al.}~\cite{davidson} give $M_{LQ}/\lambda$ limits in the range
of $1.6$ to $2.2 \ {\rm TeV}$ for scalar leptoquarks and
$2.2$ to $3.1 \ {\rm TeV}$ for vector leptoquarks,
while Leurer~\cite{leurer} arrives at values up to a factor of $\sim 1.5$
either lower or higher, and generally gets larger values for scalar than for
vector leptoquarks.
Despite of these uncertainties, our most stringent limits for vector
leptoquarks are within a factor of two close to the ones
extracted from the very low energy experiments.
They provide, however, very useful complementary information at much higher
momentum transfers from a model independent analysis involving no theoretical
assumptions on higher order corrections.

\bigskip\noindent
{\bf Compositeness Scales} \\
If quarks and leptons have a substructure and have common constituents they may
form composite objects.
Such virtual states are characterized by a compositeness scale parameter
$\Lambda$ and a coupling strength $g$, which is
set to $g^2 / 4\,\pi = 1$ in the present analysis.

The results for lower limits on the $e\,q$ compositeness scale
parameters $\Lambda^\pm$ are summarized in Table~\ref{compositeness}.
They vary between 1.0 TeV and 2.5 TeV, depending on the chiral
structure and the sign of interference with the Standard Model currents.
The bounds with positive interference are more stringent than those
with negative interference.
There is almost no difference between various lepton and quark chiralities.

\begin{table}[htb]
\begin{center}
\begin{tabular}{c c c c c c c c}
   \hdick \\[-1.5ex]
   \ $\Lambda^+_{LL}$ \ & \ $\Lambda^-_{LL}$ \ &
   \ $\Lambda^+_{LR}$ \ & \ $\Lambda^-_{LR}$ \ &
   \ $\Lambda^+_{RL}$ \ & \ $\Lambda^-_{RL}$ \ &
   \ $\Lambda^+_{RR}$ \ & \ $\Lambda^-_{RR}$ \ \\[1ex]
   \hdick \\[-1.5ex]
    2.3  & 1.0  &
    2.5  & 1.2  &
    2.5  & 1.2  &
    2.3  & 1.0  \\
\end{tabular}
\end{center}
\caption{Lower limits at 95\% confidence level on compositeness scale
         parameters $\Lambda^\pm \ [ {\rm TeV} ]$ for chiralities
         {\em LL, LR, RL} and {\em RR} with positive and negative interference
         with the Standard Model currents.
         The coupling constants are $\pm g^2/\Lambda^{\pm\,2}$ with
         the convention $g^2/4\,\pi = 1$.}
\label{compositeness}
\end{table}

Although it is common practice to give bounds on $\Lambda^\pm$, it is more
appropriate to fit directly $\eta_{if} = g^2/\Lambda^2_{if}$ as used
in the contact interaction Lagrangian.
This has the advantage that the errors behave rather gaussian.
Limits on $\Lambda^\pm$ for a certain interference are then derived from
the corresponding upper and lower values of $1/\Lambda^2$ at a given
confidence level.
A fit of $1/\Lambda^2$ yields
$\Lambda^{-2}_{LL} = \Lambda^{-2}_{RR} = (-0.41 \pm 0.33)\ {\rm TeV}^{-2}$
and $\Lambda^{-2}_{LR} = \Lambda^{-2}_{RL} = (-0.30 \pm 0.23)\ {\rm TeV}^{-2}$,
assuming a coupling strength of $g^2/4\,\pi = 1$.

The compositeness scale parameters $\Lambda$ can be interpreted in terms of a
radius of the $e\,q$ system via
$R_{e\,q}  =  \sqrt{4\,\pi / g'^{\,2}}\,\Lambda^{-1}$.
Depending on the chiral structure and the sign of interference the size
of a composite $e\,q$ state
is constrained to $R_{e\,q} \lesssim \ (0.8 \div 2)\cdot 10^{-17}\ {\rm cm}$,
if a strong coupling strength $g'$ is assumed.

Our limits on $\Lambda^\pm$ are comparable to those derived in
similar studies at $p\,\bar{p}$ and $e^+\,e^-$ col\-liders~\cite{pdg}.
For example CDF quotes $\Lambda^+_{LL} > 1.7 \ {\rm TeV}$ and
$\Lambda^-_{LL} > 2.2 \ {\rm TeV}$ for the light $u$ and $d$ quarks and
VENUS quotes $\Lambda^+_{LL} > 1.2 \ {\rm TeV}$ and
$\Lambda^-_{LL} > 1.6 \ {\rm TeV}$
assuming flavour universality for five quarks.

\bigskip\noindent
{\bf Form Factors} \\
An alternative method to study possible fermion substructures is to assign a
finite size of radius $R$ to the leptons and/or quarks~\cite{koepp}.
A convenient parametrization is to introduce in a `classical' way
form factors $f(Q^2)$ at the gauge boson--fermion vertices, which depend on the
squared momentum transfer $Q^2$
\begin{eqnarray*}
  f (Q^2) & = & 1 - \frac{1}{6}\, R^2\,Q^2 \ .
\end{eqnarray*}
For simplicity, the radius $R$ is assumed to be universal for the
electromagnetic and the weak vector and axial-vector fermion couplings.
The form factors reduce to unity and the couplings to the familiar Standard
Model  values in the pointlike limit.
A finite extension of a lepton or quark is expected to diminish the Standard
Model cross section at high $Q^2$ according to
\begin{eqnarray*}
  \frac{d\sigma}{dQ^2} & = &
  \frac{d\sigma^{SM}}{dQ^2} \, f^2_e(Q^2)\,f^2_q(Q^2) \ .
\end{eqnarray*}

The data are analyzed in terms of a single form factor, yielding as
an  upper limit at 95\% confidence level a radius of
\begin{eqnarray*}
  R & < & 2.6 \cdot 10^{-16}\ {\rm cm} \ .
\end{eqnarray*}
This result may be interpreted as a limit on the light quark sizes, since the
pointlike nature of the electron is already established down to much lower
distances in $e^+\,e^-$ and $(g - 2)_e$ experiments~\cite{kinoshita}.
Fig~\ref{xratio}~b shows the effect of a form factor with a quark radius $R_q$
given by the experimental limit on the differential cross sections;
again the sensitivity rises with $Q^2$.

The limit on $R_q$ is within a factor of two comparable to the bounds derived
from a global analysis of $Z$ decays at LEP~\cite{koepp}.

Similar upper limits on a quark radius (up to a factor of $\sqrt{6}$) are
obtained from the above contact term analysis, if the compositeness scale
parameters $\Lambda^-$ are evaluated at the electro\-magnetic scale
$g^2/4\,\pi = \alpha$.
Note, however, that in this interpretation the size is inferred from the
interference of the Standard Model currents with a new virtual current.

\section{Conclusions}
An analysis of searches for new phenomena beyond the Standard Model mediated
through contact interactions in deep inelastic
scattering at HERA has been presented.
The data correspond to a 9 fold increase in integrated luminosity compared to
a previous publication~\cite{h1lq}.

The differential cross sections $d\sigma/dQ^2$ have been measured for deep
inelastic neutral current $e^\pm\,p$ scattering in the $Q^2$ range between
$160 \ \GeV^2$ and $20,000 \ \GeV^2$.
No significant deviations from the Standard Model have been observed for
either lepton charge.

Substantially improved limits at 95\% confidence level on masses and couplings
of new heavy lepto\-quarks and on fermion compositeness scales have been
obtained, using the combined $e^-\,p$ and $e^+\,p$ cross section data.

Eight out of fourteen possible leptoquark couplings yield lower limits on
$M_{LQ}/\lambda$ which exceed the centre of mass energy of HERA,
assuming a strong coupling $\lambda = 1$, and therefore nicely complement the
searches for direct production.
Vector leptoquarks yield stronger limits than scalar leptoquarks and approach
bounds on $M_{LQ}/\lambda$ of $1 \ {\rm TeV}$.

A conceivable $e\,q$ compositeness or fermion substructure
can be ruled out for scale para\-meters
$\Lambda^\pm$ smaller than 1.0 TeV to 2.5 TeV, depending on the assumed
fermion chiralities and the sign of interference with the Standard Model
currents.

Finally, a form factor analysis constrains the size of the light
$u$ and $d$ quarks to radii smaller than $R_q < 2.6 \cdot 10^{-16}\ {\rm cm}$.

The contact interaction concept has been shown to be a very powerful tool,
becoming even more important in future high statistics data analyses.
The sensitivity to new virtual boson exchanges roughly scales as
$({\cal L}\,s)^\frac{1}{4}$ with integrated luminosity ${\cal L}$
and centre of mass energy squared $s$.

\bigskip\bigskip
\noindent
{\bf Acknowledgements.}
We are grateful to the HERA machine group, whose outstanding efforts made
this experiment possible. We appreciate the immense contributions of the
engineers and technicians, who constructed and maintained the detector.
We thank the funding agencies for financial support. We acknowledge the
support of the DESY technical staff. We also wish to thank the DESY
directorate for the hospitality extended to the non-DESY members of the
H1 collaboration.

\end{document}